\documentclass[conference]{IEEEtran}
\IEEEoverridecommandlockouts

\usepackage{cite}
\usepackage{amsmath,amssymb,amsfonts}
\usepackage{algorithmic}
\usepackage{graphicx}
\usepackage{textcomp}
\usepackage{xcolor}
\def\BibTeX{{\rm B\kern-.05em{\sc i\kern-.025em b}\kern-.08em
    T\kern-.1667em\lower.7ex\hbox{E}\kern-.125emX}}

\graphicspath{{./}}

\begin{document}

\title{Ease and Equity of Point of Interest Accessibility via Public Transit in the U.S.
\thanks{\textsuperscript{*}All authors contributed equally to this research.}
}

\author{
\IEEEauthorblockN{Alexander Li\textsuperscript{*}, Aurimas Racas\textsuperscript{*}, Junaid Syed\textsuperscript{*}, Mengyang Liu\textsuperscript{*}, Przemek Zientala\textsuperscript{*}, Tejas Santanam\textsuperscript{*}}
\IEEEauthorblockA{\textit{Georgia Institute of Technology}\\
Atlanta, GA, USA}
}

\maketitle

\begin{abstract}
Equitable access to essential Points of Interest (POIs) such as healthcare facilities and grocery stores via public transit is a critical urban planning challenge. This paper introduces an interactive tool that analyzes the ease and equity of such access in major U.S. cities. Using a 2-step floating catchment area (2SFCA) approach, we calculate granular accessibility indices and present them on an interactive, web-based platform. A key contribution of this work is a scenario analysis feature that allows users to simulate the impact of adding or removing POIs and visualize the resulting changes in accessibility equity. Our analysis reveals significant disparities across demographic groups, and we further utilize machine learning models to demonstrate a strong, quantifiable link between accessibility scores and neighborhood racial composition, highlighting systemic inequities. The tool provides a practical framework for policymakers to make more informed and equitable infrastructure decisions.
\end{abstract}


\section{Introduction}
Public transit is a cornerstone of urban life, providing essential access to employment, healthcare, food, and social opportunities, particularly for populations without access to private vehicles. Despite its importance, decision-making processes for infrastructure and Point of Interest (POI) site placement often neglect transportation accessibility and equity as primary criteria. This oversight can perpetuate and even worsen spatial inequalities within a city.

A critical issue is that transit planning itself can become "accessibility-blind." In Atlanta, for example, transit planning has historically focused on maximizing ridership counts rather than on network coverage or connectivity to essential services. This approach fails to account for areas where low ridership is a \textit{symptom} of poor connectivity to the POIs residents need to access. Consequently, a detrimental cycle emerges: areas with poor connectivity see low ridership, leading to disinvestment and further marginalization, while resources are concentrated in already well-served corridors.

To address these shortcomings, this paper presents a framework and an associated interactive tool to provide a holistic overview of public transit equity and effectiveness in major U.S. cities. Our contribution provides an interactive interface for decision-makers and the public to explore the drivers of transit network performance and identify specific gaps in accessibility. The work is designed to be extensible, using data sources that are often available globally, thus creating a potential framework for non-U.S. geographies. To demonstrate our methodology and its findings, this paper uses the city of Atlanta as a primary case study.

\section{Related Work}
The study of accessibility via public transit has become a significant area of research, motivated by the need to support populations that rely on it for essential services \cite{martin2008taking}. The COVID-19 pandemic further underscored this importance, with numerous studies demonstrating the critical role of public transit in equitable vaccine distribution \cite{thoumi2021prioritizing} and highlighting access disparities for minority communities \cite{reitsma2021quantifying} and the elderly \cite{bentivegna2022access}. Beyond the pandemic context, spatial accessibility is a key factor in accessing quality health services \cite{wan2012three}, grocery stores \cite{jiao2021measuring}, educational facilities \cite{yenisetty2020spatial}, and employment \cite{horner2005examining}. As transit agencies face post-pandemic challenges like lower ridership and service cuts \cite{chen2020transportation}, comprehensive tools to analyze and improve equitable access are more critical than ever.

A dominant methodology in this field is the Floating Catchment Area (FCA) family of techniques, which measure accessibility by relating the supply of services to the population within a given travel time. The 2-Step Floating Catchment Area (2SFCA) method is widely used for its balance of effectiveness and interpretability \cite{ghorbanzadeh2021spatial}. Researchers have also proposed several enhancements to address specific analytical needs, including the Enhanced 2SFCA (E2SFCA) to account for distance decay within catchments \cite{rader2021spatial, mohammadi2021measuring}, the 3-Step FCA (3SFCA) to better moderate for demand \cite{wan2012three, rekha2017accessibility}, and the Kernel Density 2SFCA (KD2SFCA) to smooth geographic disparities \cite{doi:10.1068/b36149}.

These methods have been applied globally to uncover significant inequities. Studies in the U.S. have used E2SFCA to identify "vaccine deserts" in areas with vulnerable populations \cite{rader2021spatial}, while research in England revealed that accounting for transit travel time uncovers disparities not visible in official statistics \cite{duffy2022evaluating}. In Brazil, an analysis using a balanced 2SFCA method found that access to intensive care units was substantially lower for Black and low-income communities \cite{pereira2021geographic}. Similarly, gravity models have shown how public transit can increase spatial inequality in accessing urban parks in Hong Kong \cite{CHANG2019111}. A common theme across these studies is that spatial access is not uniform and often disadvantages marginalized groups.

While existing work provides robust methods and consistently highlights inequities, it is often presented in static formats (e.g., reports, papers) and is typically focused on a single POI type within a single geographic area. This limits its utility for policymakers and planners who need to compare scenarios and understand complex, multi-faceted urban systems. Our work addresses this gap by developing a unified, interactive platform that allows for the analysis of multiple POI categories across several major cities. Crucially, by enabling scenario analysis, our tool transforms static accessibility metrics into a dynamic decision-support system, empowering users to explore potential interventions and their impact on transportation equity.

\section{Methodology}
Our methodology comprises three core stages: (A) data sourcing and preparation, (B) calculation of accessibility indices, and (C) development of an interactive user interface for analysis and innovation.

\subsection{Data Sources and Preparation}
\subsubsection{Data Sources}
Our analysis integrates four primary data categories from five major U.S. metropolitan areas (New York City, Los Angeles, Chicago, Dallas, and Atlanta):
\begin{itemize}
    \item \textbf{Demographic Data:} Population counts, racial and ethnic breakdowns, income brackets, and vehicle availability at the census block group level, sourced from the 2020 American Community Survey (ACS) via SafeGraph.
    \item \textbf{Point of Interest (POI) Data:} Locations and names for schools, theaters, restaurants, hospitals, and grocery stores, collected from OpenStreetMap (OSM) using the Overpass API. This was supplemented with vaccination site data from the Centers for Disease Control (CDC). Our final dataset included approximately 60,000 vaccination centers, 14,000 restaurants, 4,500 grocery stores, 4,200 schools, and 700 hospitals.
    \item \textbf{Public Transit Data:} Service schedules and stop information from General Transit Feed Specification (GTFS) feeds provided by each city's public transit operators.
\end{itemize}

\subsubsection{Data Preparation}
\textbf{Catchment Area Estimation:} For each POI, we estimated its service area by generating 30-minute public transit isochrones—a boundary delineating the area reachable within that travel time. These were calculated for morning, afternoon, and evening schedules using a containerized OpenTripPlanner server, which leverages GTFS and OSM data to produce accurate, time-dependent travel polygons.

\textbf{Hierarchical Spatial Indexing:} To ensure consistent spatial units for analysis and visualization across all cities, we mapped all data onto the H3 hierarchical spatial indexing system. We used H3 resolution 9, which corresponds to hexagons with an average area of approximately 0.1~km². Demographic data from irregular census block group polygons was apportioned to the hexagons they overlap with, proportional to the area of intersection. This process standardizes the geography and mitigates issues arising from the varying size and shape of census administrative boundaries.

\begin{figure}[htbp]
    \centering
    \includegraphics[width=3in]{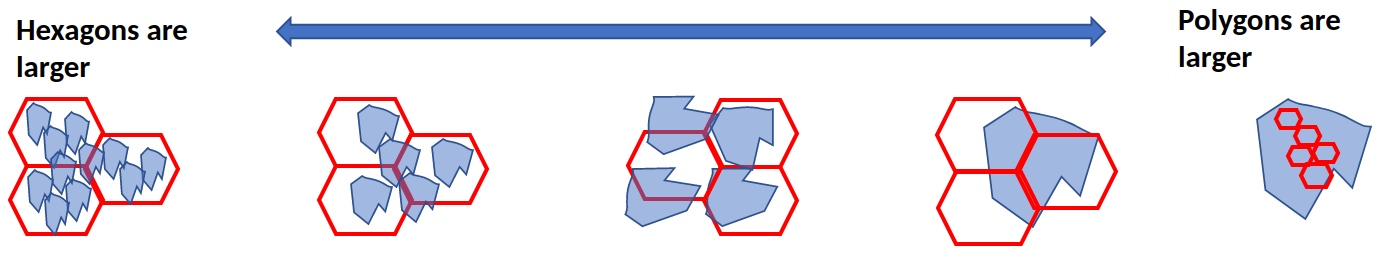}
    \caption{Illustration of the spatial discretization process. Irregular census block group polygons (blue) are mapped onto a standardized H3 hexagonal grid (red). Demographic data is apportioned based on area overlap, creating a consistent spatial unit for analysis.}
    \label{fig:h3mapping}
\end{figure}

\subsection{Accessibility Index Calculation}
We selected the 2-Step Floating Catchment Area (2SFCA) method \cite{ghorbanzadeh2021spatial} to calculate the accessibility index. This method was chosen for its strong balance of analytical robustness and intuitive simplicity, making its results accessible to non-technical stakeholders such as city planners. The calculation proceeds in two steps:

\begin{enumerate}
    \item \textbf{Step 1: Calculate POI-to-Population Ratio.} For each POI $j$, we compute its service ratio, $R_j$, by dividing its supply (assumed to be 1 for each location) by the total population within its 30-minute transit catchment area, $C_j$.
    \begin{equation}
        R_j = \frac{S_j}{\sum_{k \in C_j} P_k}
    \end{equation}
    where $S_j$ is the supply of POI $j$ (here, $S_j=1$), and $P_k$ is the population of a given hexagon $k$ within the catchment $C_j$.
    
    \item \textbf{Step 2: Sum Ratios for Each Location.} For each hexagon $h$, we calculate its final accessibility score, $A_h$, by summing the ratios $R_j$ of all POIs whose catchment areas cover that hexagon.
    \begin{equation}
        A_h = \sum_{j: h \in C_j} R_j
    \end{equation}
\end{enumerate}

\subsection{User Interface and Innovations}
\begin{figure}[htbp]
    \centering
    \includegraphics[width=3.5in]{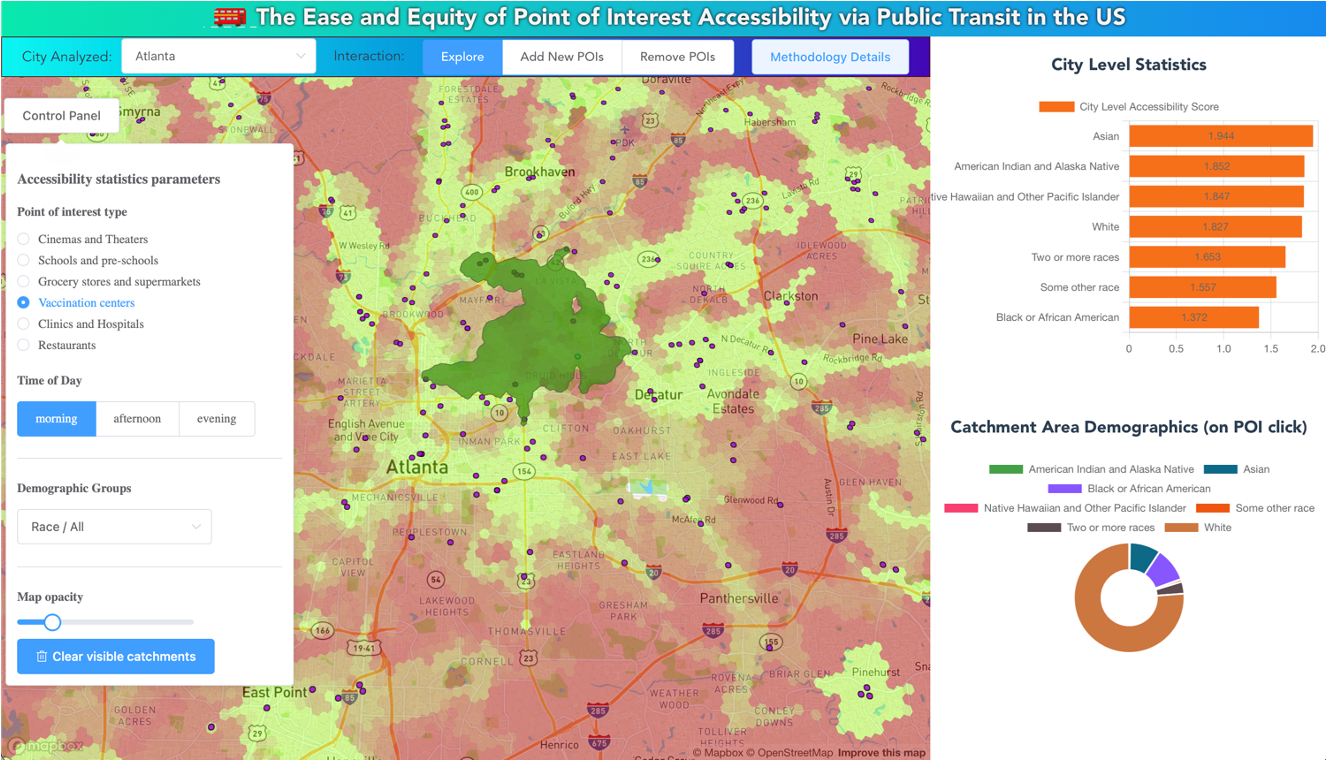}
    \caption{Screenshot of the interactive web-based user interface. The main panel displays a choropleth map of accessibility scores in Atlanta. The left control panel allows users to select POI type, city, and demographic filters, while charts on the right display aggregate statistics.}
    \label{fig:ui}
\end{figure}

A primary contribution of this work is an interactive, web-based tool built with Vue.js, Deck.gl, and a FastAPI backend. The tool (Fig. \ref{fig:ui}) empowers users to explore accessibility dynamics in several novel ways.

The interface allows users to filter the analysis by city, POI type, time of day, and various demographic factors (race, income, vehicle availability). This reveals how accessibility differs for specific populations. The key innovation is the \textbf{scenario analysis} feature: users can hypothetically add or remove POIs on the map, and the tool dynamically recalculates all accessibility scores and equity metrics. This transforms the analysis from a static report into a dynamic decision-support system.

Furthermore, our approach introduces several innovations over existing work. The use of the H3 grid provides a standardized geography, enabling direct comparison between different parts of a city or even across different cities. The granularity of our analysis, down to the census block group level, provides a highly accurate view of equity that is often lost in analyses performed at coarser levels like the metropolitan statistical area (MSA). Finally, the combination of multi-city, multi-POI data with an interactive scenario analysis tool represents a novel platform for applied accessibility research.

\section{Results and Discussion}
Our evaluation consists of two parts. First, we present a comparative analysis of accessibility scores across cities and demographic groups. Second, we employ statistical models to formally quantify the extent of the observed inequities.

\subsection{Comparative and Equity Analysis}
\begin{figure}[htbp]
    \centering
    \includegraphics[width=3.5in]{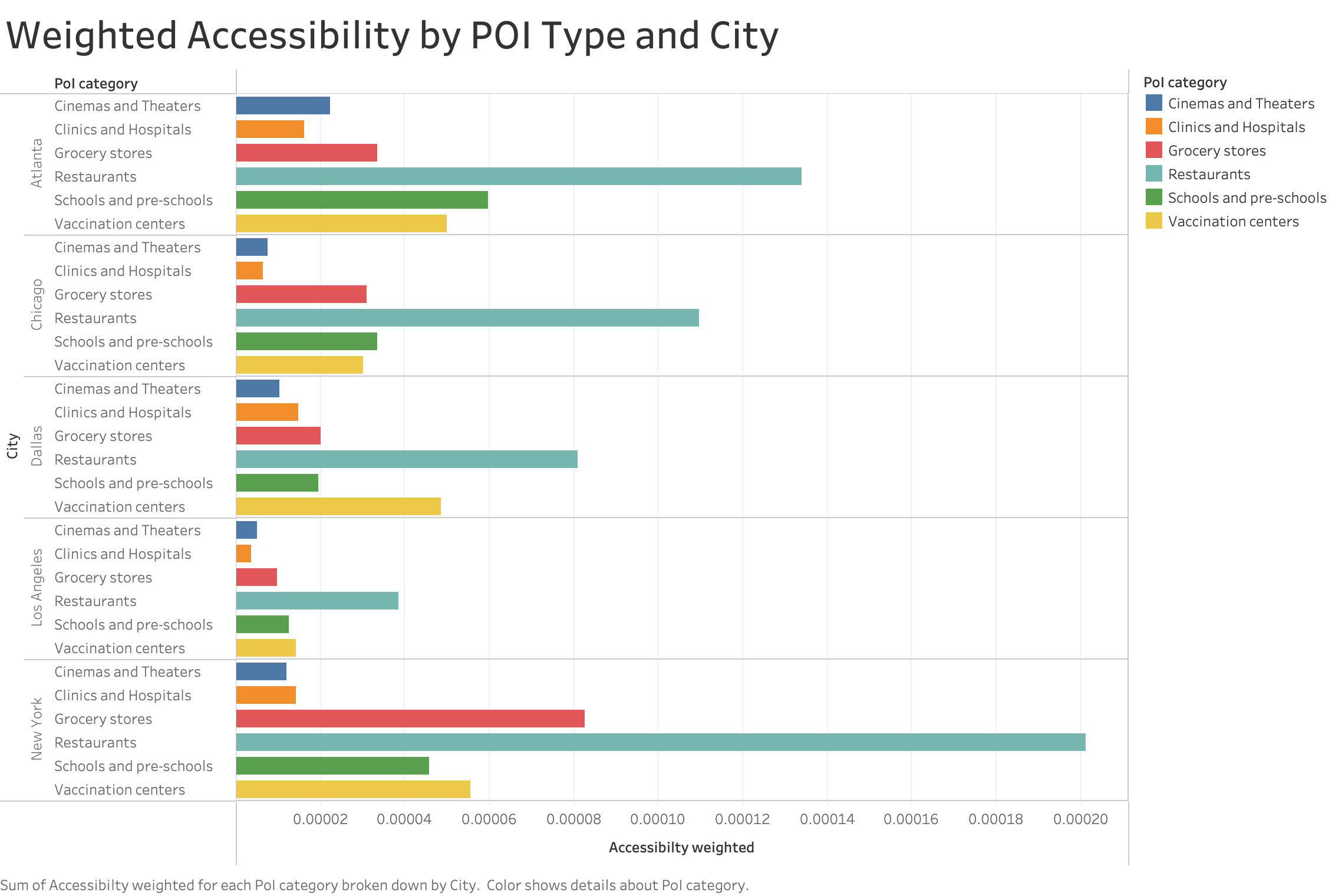}
    \caption{Population-weighted accessibility scores by POI type across five major U.S. cities. Higher scores indicate better access. New York City generally leads, while Los Angeles consistently lags.}
    \label{fig:accessbypoi}
\end{figure}

Our cross-city comparison of population-weighted accessibility scores reveals significant performance differences among urban transit systems (Fig. \ref{fig:accessbypoi}). New York City consistently demonstrates the highest accessibility for POIs like grocery stores, restaurants, and vaccination centers, a finding that aligns with its high population and service density. Conversely, Los Angeles ranks lowest across nearly all POI categories, reflecting its notoriously challenged public transit system. Atlanta presents a mixed but relatively strong profile, with the highest accessibility for theaters, hospitals, and schools among the cities analyzed.

\begin{figure}[htbp]
    \centering
    \includegraphics[width=3.5in]{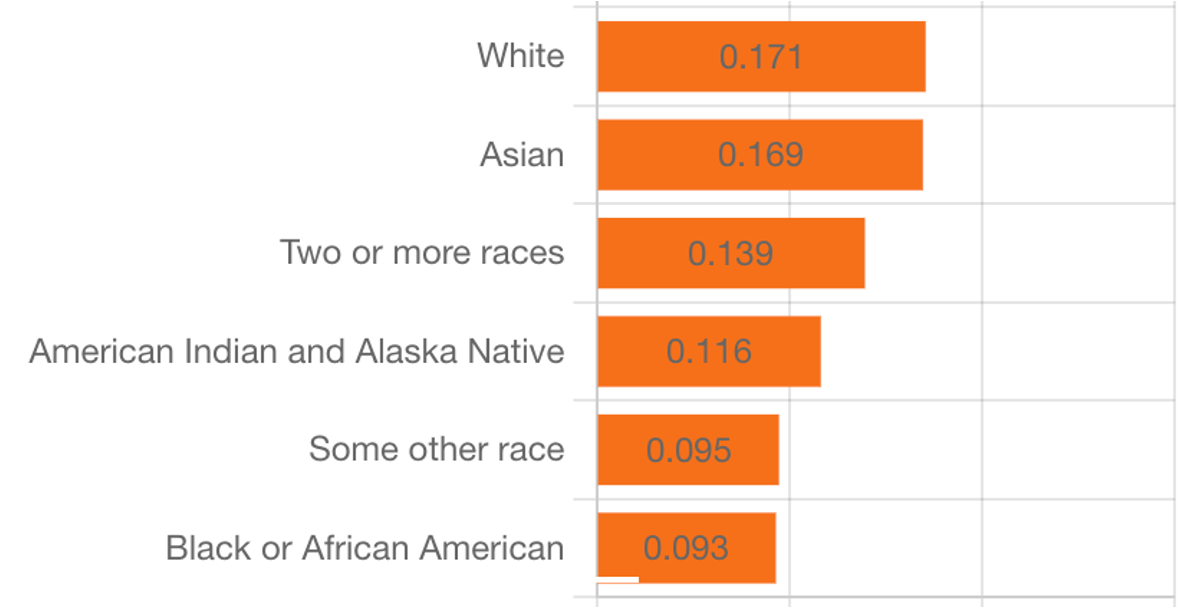}
    \caption{Disparities in hospital accessibility by race in Atlanta. The average accessibility index for the White population (0.171) is substantially higher than for the Black or African American population (0.093), indicating severe inequity.}
    \label{fig:hospitalstats}
\end{figure}

However, a high overall accessibility score can mask significant internal inequities. In Atlanta, despite its high overall ranking for hospital access, our analysis reveals a stark racial divide. As shown in Fig. \ref{fig:hospitalstats}, the average hospital accessibility index for White populations is nearly double that of Black populations. A similar disparity exists for grocery stores, where access is 70\% higher for White populations than for Black populations.

In contrast, other cities show different patterns. In Chicago, the accessibility gap for grocery stores between White and Black populations is much smaller, and for hospitals, Black populations exhibit slightly higher accessibility than White populations. Even Los Angeles, despite its poor overall performance, shows more equitable access across racial lines for most POIs. This suggests that while the L.A. transit system is broadly suboptimal, it does not disadvantage specific racial groups as severely as Atlanta's system does. These findings for Atlanta point to a systemic issue where the public transit authority (MARTA) provides insufficient coverage in the majority-Black South and Southwest portions of the city, creating a deep-seated transit divide.

\subsection{Statistical Modeling of Inequity}
To move beyond descriptive statistics and formally quantify the relationship between geography, race, and access, we conducted two machine learning experiments.

\begin{figure}[htbp]
    \centering
    \includegraphics[width=3.5in]{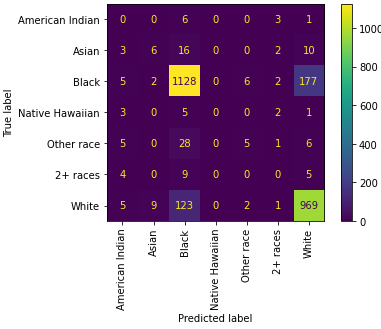}
    \caption{Confusion matrix for a model predicting majority race from accessibility scores. The model's ability to accurately distinguish between hexagons with Black and White majorities underscores the strong correlation between race and transit access.}
    \label{fig:confmat}
\end{figure}

First, we trained a Random Forest classifier to predict the majority race of an H3 hexagon using only its accessibility score to vaccination centers. The model achieved high predictive certainty for hexagons with White or Black majorities (F1-scores $\approx 0.85$), but performed poorly for other minority races due to their smaller sample sizes. The confusion matrix for the test set (Fig. \ref{fig:confmat}) illustrates this high predictive power. The fact that race can be accurately predicted from a transit accessibility metric alone is a stark statistical indicator of systemic inequity and residential segregation.

Second, we analyzed the dynamic from the opposite direction by training an Extra Trees Regressor to predict a hexagon's accessibility score based on its demographic characteristics. The model performed well (RMSE of 0.2), and a feature importance analysis revealed that the most predictive characteristics, in order, were race, age/sex composition, vehicle availability, and income. This confirms that demographic factors, particularly race, are primary drivers of the observed accessibility patterns.

\section{Conclusion}
This paper presented an interactive, data-driven tool designed to empower city planners, researchers, and the public to analyze public transit accessibility and equity. By integrating demographic data with a robust 2SFCA accessibility metric and a dynamic scenario-analysis feature, our work provides a powerful lens through which to view urban inequality. The implications of such analysis are profound, particularly for access to critical services like healthcare, where equitable transit can be a life-or-death matter \cite{who}.

Our principal findings reveal stark disparities both between and within major U.S. cities. While some cities like Los Angeles suffer from low overall accessibility, others like Atlanta exhibit a paradox of high aggregate scores that mask deep-seated racial inequities. The finding that hospital accessibility for White populations in Atlanta is nearly double that for Black populations highlights the urgent need for transit planning to evolve beyond simple ridership metrics. Instead, it must adopt an equity-focused approach to ensure that underserved populations receive the vital transportation connections necessary for their livelihood.

The primary limitations of this study are twofold: the potential for incomplete POI datasets and the exclusive focus on public transit. While we utilized the best available open data, POI locations from sources like OSM may not be exhaustive. Furthermore, our analysis does not account for other transport modes like driving, walking, or cycling. However, by establishing a baseline for accessibility via public transit, a critical mode for many vulnerable populations, our work provides an essential check on whether services are reachable by all. Future research could address these limitations by integrating proprietary or governmental POI data and expanding the model to incorporate multi-modal transport networks.

Ultimately, this work demonstrates that the tools and data exist to quantify and visualize complex urban inequities that have long been discussed qualitatively. By making these dynamics transparent and interactive, we hope to foster data-informed dialogue and drive policy changes that lead to more just, accessible, and equitable cities for all residents.


\end{document}